\documentclass[twocolumn,prc,showpacs,preprintnumbers,superscriptaddress]{revtex4-2}

\usepackage[T1]{fontenc}
\usepackage{amsmath} 
\usepackage{amssymb}
\usepackage{amsfonts}
\usepackage{mathrsfs}
\usepackage{units}
\usepackage{multirow}
\usepackage{color}
\usepackage{braket}

\usepackage{dcolumn}
\usepackage{bm}
\usepackage{rotating}

\newcommand{\disregard}[1]{}

\newcommand{\gras}[1]{\boldsymbol{#1}}

\usepackage{tikz}

\newsavebox{\tmpstrikebox}
\newlength{\tmpstrikelen}

\graphicspath{{Figures/}, {Tables/}}

\begin{document}

\title{Mirror energy differences in $\bm{T=1/2}$ $\bm{f_{7/2}}$-shell nuclei within isospin-dependent 
DFT}

\author{P. B\k{a}czyk}
\affiliation{Institute of Theoretical Physics, Faculty of Physics, University of Warsaw, ul. Pasteura 5,
PL-02-093 Warsaw, Poland}

\author{W. Satu{\l}a}
\affiliation{Institute of Theoretical Physics, Faculty of Physics, University of Warsaw, ul. Pasteura 5,
PL-02-093 Warsaw, Poland}

\date{\today}

\begin{abstract}
\begin{description}
\item[Background]
         Small asymmetry between neutrons and protons, caused by the  differences in masses and 
charges of the up and down constituent quarks, leads to isospin symmetry breaking.
The isospin non-conservation affects a broad range of observables from superallowed Fermi weak interaction to isospin-forbidden electromagnetic rates. Its most profound and cleanest manifestation are systematic shifts in masses and excitation energies of mirror atomic nuclei. 
\item[Purpose]
         Recently, we constructed the  charge-dependent density functional theory (DFT)  that includes class~II and III local interactions and demonstrated that the model allows for very accurate reproduction of Mirror and Triplet Displacement energies  in a very broad range of masses. The aim of this work is to  further test the charge-dependent functional by studying Mirror Energy Differences (MEDs) in the function of angular momentum $I$. 
\item[Methods]
         To compute MEDs we use a DFT-rooted no core configuration interaction model. This  post
         mean-field method  restores rotational symmetry and takes into account configuration mixing within a space that includes relevant  (multi)particle-(multi)hole Slater determinants.
\item[Results]
         We applied the model to $f_{7/2}$-shell mirror pairs of $A$=43, 45, 47, and 49 
         focusing on MEDs in the low-spin part (below band crossing), which allowed us to limit the model space to seniority one and three (one broken pair) configurations.          
\item[Conclusions]
         We demonstrate that, for spins $I\leq 15/2$ being the subject of the present study, our model 
reproduces well experimental MEDs which vary strongly in the function of $I$ and $A$.  
The quality of the model's  predictions for MEDs is comparable to the nuclear shell-model results by Bentley {\it et al.\/} 
Phys.\ Rev.\ C {\bf 92}, 024310 (2015).

\end{description}
\end{abstract}

\maketitle

\section{Introduction}\label{intro}

Isospin symmetry breaking (ISB) in finite nuclei reflects a subtle balance 
between the isospin symmetry violating long-range electrostatic interaction that 
polarizes the entire nucleus and the short-range strong force, which, predominantly, preserves the  
isospin symmetry. These two scales are intertwined, which means that the accurate theoretical treatment 
of the ISB effects is a highly nontrivial task. 

Mean-field or Single Reference Energy Density Functional (SR-EDF)-based methods are essentially 
the only techniques that allow proper treatment of the long-range polarization effects, in a fully self-consistent 
fashion, over the entire nuclear chart.  Moreover, because these methods use effective  short-range interactions that are constructed using low-$q$ expansion they allow
for systematic inclusion of the  isoscalar as well as the ISB short-range forces on the same footing. The later components are indispensable not only in reproducing Mirror (MDEs) and Triplet (TDEs) Displacement Energies, 
the primary isovector and isotensor observables, see~\cite{(Orm89),(Bro00),(Bac18),(Bac19)}, but also in calculating properties of isobaric analog states (IAS) in heavy nuclei like $^{208}$Pb or $^{208}$Bi~\cite{(Col98), (Roc18)}.

Recently, we developed SR-DFT that includes, apart of the Coulomb interaction, the generalized Skyrme force 
consisting  of a standard isoscalar Skyrme part, the leading-order (LO) zero-range~\cite{(Bac18)} and next-to-leading order (NLO) gradient interactions~\cite{(Bac19)} of class~II and~III in the Henley and Miller classification~\cite{(Hen79),(Mil95)}.  
These forces introduce  charge-independence (CIB) and charge-symmetry breaking (CSB) short-range effects, respectively. With these terms we were able to reproduce very accurately almost all, except for the very few lightest $A<6$ cases, existing data on MDEs and TDEs~\cite{(Bac18),(Bac19)}.  In Ref.~\cite{(Bac19)} we have also provided the arguments that the newly introduced ISB terms model strong-force-related effects of CIB and CSB rather than the beyond-mean-field electromagnetic corrections.
The aim of this work is to  test the consistency of our generalized charge-dependent EDF in the CSB channel by computing  Mirror Energy Differences (MEDs) in rotational bands  of $A$=43, 45, 47, and 49 $T$=1/2 mirror nuclei at low-spins. The MEDs are defined as follows:
\begin{equation}
{\rm MED}(I) = \Delta E_{I,T, -T_z} - \Delta E_{I, T, T_z}
\end{equation}
where $\Delta E_{I, T, \pm T_z}$ is the excitation energy of a state of given spin $I$ and isospin $T$ in 
a nucleus  with $ \pm T_z$.
The low-spin MEDs in these nuclei data vary strongly with $A$ thus posing a challenging task for the theory.    

Until very recently, such calculations were reserved almost exclusively for the Nuclear Shell Model (NSM), a configuration-interaction (CI) approach involving strict laboratory-frame treatment of symmetries. The NSM prescription for MEDs was formulated in Ref.~\cite{(Zuc02)} and  subsequently applied to $sd$- and $fp$-shell 
nuclei in  Refs.~\cite{(Ben07),(Kan13),(Kan14),(Ben15)} and references quoted therein. Mirror and Triplet Energy Differences for $A=70$ were also investigated in beyond-mean-field VAMPIR code~\cite{(Pet15)}. 
Recently, such calculations became also within the reach of symmetry-projected multi-reference DFT (MR-DFT) and its No-Core Configuration-Interaction (DFT-NCCI) extension,  see~\cite{(She19)} and references quoted therein. Our group has developed a DFT-NCCI variant involving an unpaired Skyrme functional and a 
unique combination of angular-momentum and isospin projections and applied it to calculate the spectra and beta-decay rates in  $N\approx Z$ nuclei~\cite{(Sat14d),(Sat16d),(Kon16)}.  Recently, we have incorporated into the DFT-NCCI framework the CSB contact terms~\cite{(Kon19)} and 
applied it to calculate the ISB corrections to the Fermi matrix elements in $sd$-shell $T$=1/2 mirror nuclei.
In the subsequent work~\cite{(Lle20)} we performed a seminal calculation of MEDs in the heaviest 
mirror pair measured so far $^{79}$Zr/$^{79}$Y. The results obtained so far are very promising. In particular, 
they indicate that a relatively limited number of configurations is needed to obtain a good description
of a low-energy,  low-spin physics in complex nuclei.

In this work we encroach with the DFT-NCCI method into a territory which is traditionally reserved 
for the NSM.  However, it is not our intention to suggest that the DFT-NCCI technique is an alternative 
to the NSM.  Without any doubt, the NSM is better optimized to address fine details of nuclear structure 
in the traditional regions of its applicability. It faces, however, natural computational limits hampering its 
ability to treat, for example,  heavier nuclei which, on the other hand, can be easily addressed using DFT-based techniques. 
In this sense, the DFT-NCCI model presented here should be viewed as  a complementary theoretical tool to the NSM.  
We want to also underline that the interaction used here to generate the energy density functional
is an empirical effective interaction  with low-energy coupling constants (LECs) adjusted to empirical data,
not derived from fundamental theory of nuclear interactions. In this sense our DFT-NCCI model serves as 
useful tool to compute different observables which can be compared to their {\it ab initio\/} counterparts
but only barely allows us to trace back the physics origin of these effects to the  specific properties of fundamental
nuclear interaction.

This paper is organized as follows. In Sec.~\ref{sec:ncci} we briefly overview the
charge-dependent DFT-NCCI model  paying special attention to the concept of configuration and model spaces.
In Sec.~\ref{sec:MEDs} we discuss in detail the results obtained for $A=$43, 45, 47, and 49 mirror doublets.  Summary and conclusions are presented in Sec.~\ref{sec:conc}.

\section{The DFT-NCCI model}\label{sec:ncci}

The DFT-NCCI is a post Hartree-Fock(-Bogliubov) framework, which mixes many-body states 
projected from deformed independent particle-hole or (quasi)particle configurations. 
As already mentioned, our group has developed the DFT-NCCI variant based on the unpaired Skyrme 
functional and a combination of the angular-momentum and isospin projections. 
The smallness of isospin mixing~\cite{(Sat09)} allows us to assume that the 
rigorous treatment of isospin, which is critical for isospin-breaking corrections to superallowed beta decays~\cite{(Sat11)}, should have a minor influence on the calculated spectra and MEDs.  Hence, in order to facilitate calculations, we decided 
to use  here a variant involving only angular-momentum projection.

The method proceeds as follows. Firstly, we construct a {\it configuration space\/} 
by computing self-consistently a set of physically relevant  (multi)particle-(multi)hole 
Hartree-Fock Slater determinants  $\{ \ket{\varphi_j}\}_{j=1}^{N_{\rm conf}}$.
In the next step, we build a {\it model space\/} which is composed of good
angular-momentum states projected from the mean-field configurations 
$\{ \ket{\varphi_j}\}_{j=1}^{N_{\rm conf}}$:
\begin{equation}
\ket{\varphi_j ; I M; T_z}^{(i)}=\frac{1}{\sqrt{\mathcal{N}^{(i)}_{\varphi_j ; IM;T_z}}}
\sum_{\substack{K }} a_{K}^{(i)} \hat{P}^{I}_{MK}\ket{\varphi_j} \,,  \label{eq:mrdftstate}
\end{equation} 
where $K$ stands for a projection of angular momentum onto the intrinsic $z$-axis while  
\begin{eqnarray}
\hat P^I_{M K} & = & \frac{2I+1}{8\pi^2 } \int d\Omega\;
\; D^{I\, *}_{M K}(\Omega )\; e^{-i\gamma \hat{J}_z}
e^{-i\beta \hat{J}_y} e^{-i\alpha \hat{J}_z} ,
\end{eqnarray}
is the standard angular-momentum projection 
operator. The index $i$ enumerates different states of a given spin $I$,
$\mathcal{N}^{(i)}_{\varphi_j;IM;T_z}$ is a normalization constant while
$D^{I}_{M K}(\Omega )$ is the Wigner function. The integration over the Euler angles 
$\Omega = (\alpha, \beta, \gamma )$  is performed using the Gauss-Chebyshev (over $\alpha$ and
$\gamma$) and Gauss-Legendre (over $\beta$) quadratures  
with $n_\alpha=n_\beta=n_\gamma=40$ knots to assure 
precise integration over the Euler angles for all spin states considered in this work,  
see~\cite{(Dob09d)} for further details.

The angular-momentum projected states (\ref{eq:mrdftstate}) are, in general, non-orthogonal  to each other, 
often leading to an overcomplete {\it model space\/}.
Hence, in the final step, we perform configuration-interaction calculation 
by solving the Hill-Wheeler-Griffin equation~\cite{(RS80)}. In the mixing calculation 
we use the same Hamiltonian that was used to generate the
configurations.  In effect, we obtain a set of linearly independent 
DFT-NCCI eigenstates of the form 
\begin{equation}
\ket{\psi_{\textrm{NCCI}}^{k; IM; T_z}}
=\frac{1}{\sqrt{\mathcal{N}^{(k)}_{IM;T_z}}}
\sum_{ij}c_{ij}^{(k)}\ket{\varphi_j;I M;T_z}^{(i)} \,, \label{eq:nccistate}
\end{equation}  
together with the corresponding energy spectrum. 
More details concerning our method can be found in Ref.~\cite{(Sat16d)}.

The angular-momentum projection is handled by using the generalized Wick's theorem (GWT), which 
is the only  technique that makes the method numerically tractable.  It leads, however, 
to singular kernels once modern density-dependent Skyrme or Gogny forces are used for the beyond-mean-field part of the calculation. 
In spite of many efforts to regularize the singularities~\cite{(Ben09),(Sat14b)} 
no satisfactory solution has been found so far.  Hence, at present, the theory can be safely carried on only 
for true interactions like the SLyMR0 \cite{(Sad13)} or SV$_{\rm T}$ \cite{(Bei75), (Sat14b)} density-independent 
Skyrme pseudopotentials.  In this work we shall use the SV$^{\rm ISB}_{\rm T;\, NLO}$ Skyrme pseudopotential
augmented with class~III CSB interaction (class~II force is inactive in isospin doublets):
\begin{eqnarray}
\hat{V}^{\rm{III}}(i,j)  =  \bigg[&&
t_0^{\rm{III}}  \delta\left(\gras{r}_{ij} \right)
  +  \frac12 t_1^{\rm{III}}
\left( \delta\left( \gras{r}_{ij} \right) \bm{k}^2 + \bm{k}'^2 \delta\left(\gras{r}_{ij} \right) \right)   
\nonumber \\
& &+  t_2^{\rm{III}}
\bm{k}' \delta\left(\gras{r}_{ij} \right) \bm{k} \bigg]  \left( \hat{\tau}_3^{(i)}+\hat{\tau}_3^{(j)} \right) ,
\label{eq:classIII_NLO}
\end{eqnarray}
where $\gras{r}_{ij} = \gras{r}_i - \gras{r}_j$,
$\bm{k}  =  \frac{1}{2i}\left(\bm{\nabla}_i-\bm{\nabla}_j\right)$ and
$\bm{k}' = -\frac{1}{2i}\left(\bm{\nabla}_i-\bm{\nabla}_j\right)$ are the standard relative-momentum
operators acting to the right and left, respectively.
The three new  
low-energy coupling constants   $t_0^{\rm{III}}= {}$\mbox{$11\pm2$\,MeV\,fm$^3$},
   $t_1^{\rm{III}}= {}$\mbox{$-14\pm4$\,MeV\,fm$^5$}, and $t_2^{\rm{III}} = {}$\mbox{$-7.8\pm0.8$\,MeV\,fm$^5$} 
have been adjusted to all available data on MDEs for $A\geq 6$ in Ref.~\cite{(Bac19)}. 
In this sense our approach is free from adjustable parameters.

\section{MEDs in the lower $\bm{fp}$-shell mirror doublets}\label{sec:MEDs}

At variance with NSM, the configuration and model spaces of our DFT-NCCI approach are not fixed.
In practice, we build the configuration space step-by-step by adding physically relevant low-lying 
(multi)particle-(multi)hole mean-field configurations which, in the
present calculation,  are self-consistent HF solutions conserving parity and signature symmetries.  
The basic strategy  is to explore configurations built upon all relevant single-particle (s.p.) deformed Nilsson levels   $|N n_z \Lambda\, K ; r\rangle$ where $r=\pm i$ is a quantum number associated with  the signature-symmetry  operator $\hat R_y= e^{-i\pi \hat J_y}$. 
In the lower $fp$-shell nuclei the active Nilsson orbitals that determine physics of low-spin states
are $|330\: 1/2; \pm i\rangle$, $|321\: 3/2; \pm i\rangle$, $|312\: 5/2; \pm i\rangle$, and 
$|303\: 7/2; \pm i\rangle$ originating from the spherical $f_{7/2}$-shell. Hence, in the following,
we will explore configurations  involving only these four orbitals, with the exception
of a single configuration in  the $A=$47 doublet that would involve the $|321\: 1/2 \rangle$ Nilsson orbital.

It transpired {\it a posteriori\/}  that all the calculated configurations are axial. For the mixing calculation  
we fix the orientation of the nucleus  with its symmetry axis along the $Oy$-axis. This 
allows us to associate uniquely the s.p.\ level's signature quantum number $r$ with its $K$ quantum number (along the symmetry axis) through the relation $r= e^{-i\pi K}$.  In turn, the active Nilsson 
levels can be uniquely labeled by providing the $K$ quantum number and, if needed, 
the isospin subscript $\tau = \nu$ ($\pi$) to differentiate between neutron (proton) levels, respectively. 
Note that the $K$ and $-K$ (denoted below by $\bar K$) Nilsson levels correspond to opposite signatures.  Axial symmetry implies that the total angular-momentum
projection onto the $Oy$ axis of the intrinsic system, $\Omega = \sum_{i}^{occup} K_i$, is 
conserved.  Moreover,  the signature reversed configurations are equivalent.  

The calculated excited configurations can be divided into three groups. 
The first group  includes the simplest excited configurations which 
are  p-h seniority-one ($\nu =1$) excitations involving unpaired proton or neutron. 
The strategy used in this work is to include all such configurations within the active space.  
The second group involves selected $nn$ or $pp$ seniority-zero pairing excitations and $np-$pairing excitations coupled to $\Omega = 0$ i.e. $|K_\tau \bar{K}_{-\tau}\rangle$.  
The third group involves (typically the four lowest) seniority-three  ($\nu = 3$) broken-pair configurations.
 Note that the configuration space constructed in this way neither includes multi-broken-pair configurations nor (near-)fully-aligned configurations which determine physics in the crossing region and nearby the terminating state, respectively, which restricts the present analysis to the low-spin data.  

All calculations presented in this work were done using a developing version of the HFODD solver~\cite{(Dob09d),(Sch17)} equipped with the DFT-NCCI module and including  
CSB EDF in the projection module. In the calculations  we use the basis composed of 12 spherical HO shells.  
 In order to study a sensitivity of our results
  to the short-range NLO CSB interaction we shall perform two variants of the DFT-NCCI calculations:  
  the full variant that includes both the Coulomb and short-range NLO CSB terms (termed DFT-NCCI-NLO) 
  and the variant that includes the Coulomb interaction as the only source of ISB (called DFT-NCCI-COU). 
 In both variants the exchange term of the Coulomb interaction is treated exactly.

\subsection{MEDs in the A=47 mirror pair}\label{sec:A47}

We shall start the discussion somewhat unconventionally with the case of the $A=47$ mirror pair.
The reason is that this case was  most thoroughly studied, with
the largest number of 20 configurations included in the mixing calculation. In turn,
it allows us to reason that our calculated MEDs  are well converged at low-spins and draw
conclusions concerning the impact of specific groups of configurations on the values of MEDs. 
These conclusions will be used in the remaining cases to constrain the configuration space. 

\begin{figure}[h!]
\centering
\caption{(Color online)  
Configurations used in the DFT-NCCI calculations for $A$=47. Full dots denote pairwise occupied 
 Nilsson states.  Up (down) arrows denote singly occupied 
Nilsson states with positive (negative) $K$ quantum numbers, respectively.}
\label{fig:A47CONF}
\includegraphics[width=\columnwidth]{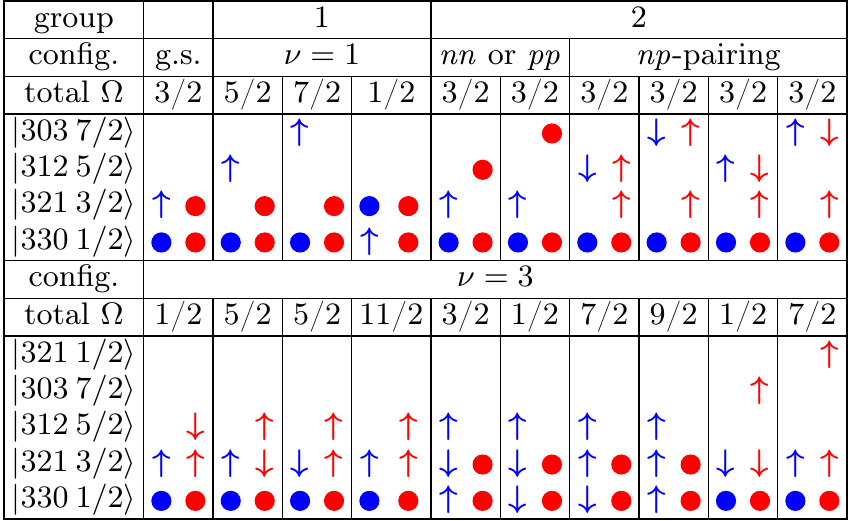}
\end{figure}

The configuration space for the $A=47$ mirror pair is schematically depicted 
in Table~\ref{fig:A47CONF}.  It includes the ground-state 
(GS) and 19 excited  (multi)particle-(multi)hole configurations. 
The first group of excitations includes the  p-h seniority-one ($\nu =1$) excitations in the {\it active\/} space. In the case of $A=47$ mirrors there are three such configurations 
corresponding to: $|3/2\rangle\rightarrow |5/2\rangle$, $|3/2\rangle\rightarrow |7/2\rangle$, and
$|\overline{1/2}\rangle\rightarrow |\overline{3/2}\rangle$ p-h excitations in the
odd-particle-number subsystem.  The second group  involves two $nn$- or $pp$-pairing excitations and four $np$-pairing excitations coupled to $\Omega = 0$ i.e. $|K_\tau \bar{K}_{-\tau}\rangle$.   
Eventually, in the third group, 
we include ten seniority-three  ($\nu = 3$) broken-pair configurations. 

\begin{figure}[thb]
\centering
\includegraphics[width=\columnwidth]{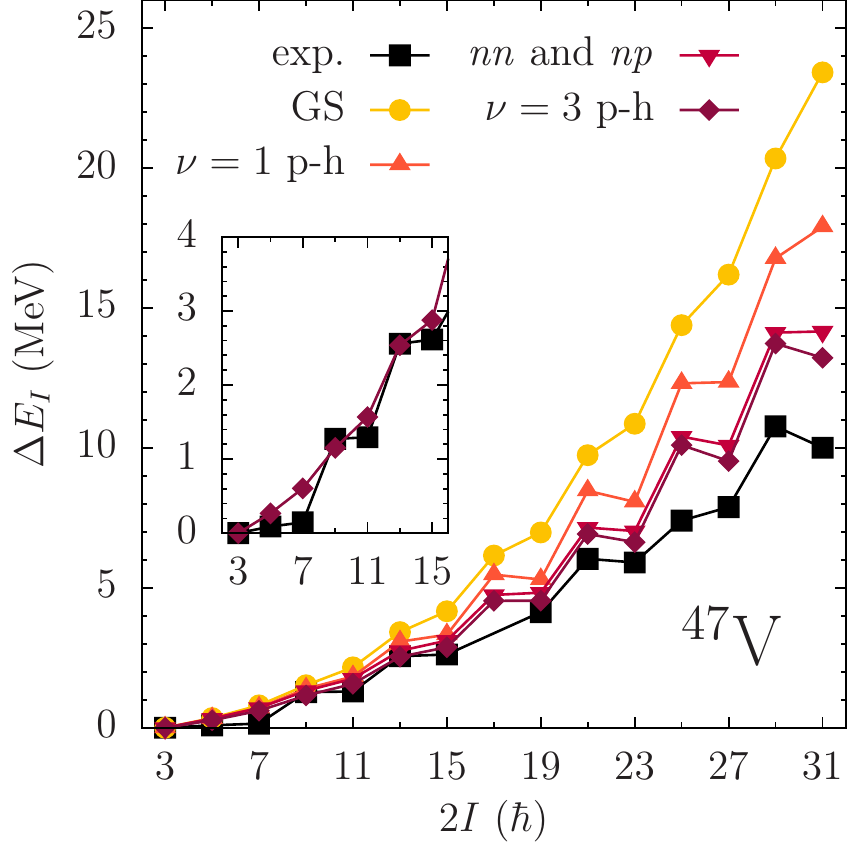}
\caption{(Color online)  
Excitation energy $\Delta E$ versus spin $I$ in the $^{47}$Cr. Squares show empirical data
taken from~\cite{(Bur07)}.
Circles mark theoretical results obtained for the GS configuration only. Triangles up 
include  $\nu =1$ p-h configurations. Triangles down take into account also 
$nn$ and $np$ pairing configurations. Diamonds mark the DFT-NCCI results including 
all configurations shown in Table~\ref{fig:A47CONF}.
The inset magnifies the low-spin part of the {\it yrast\/} spectrum.}
\label{fig:A47E}
\end{figure}

\begin{figure}[thb]
\centering
\includegraphics[width=\columnwidth]{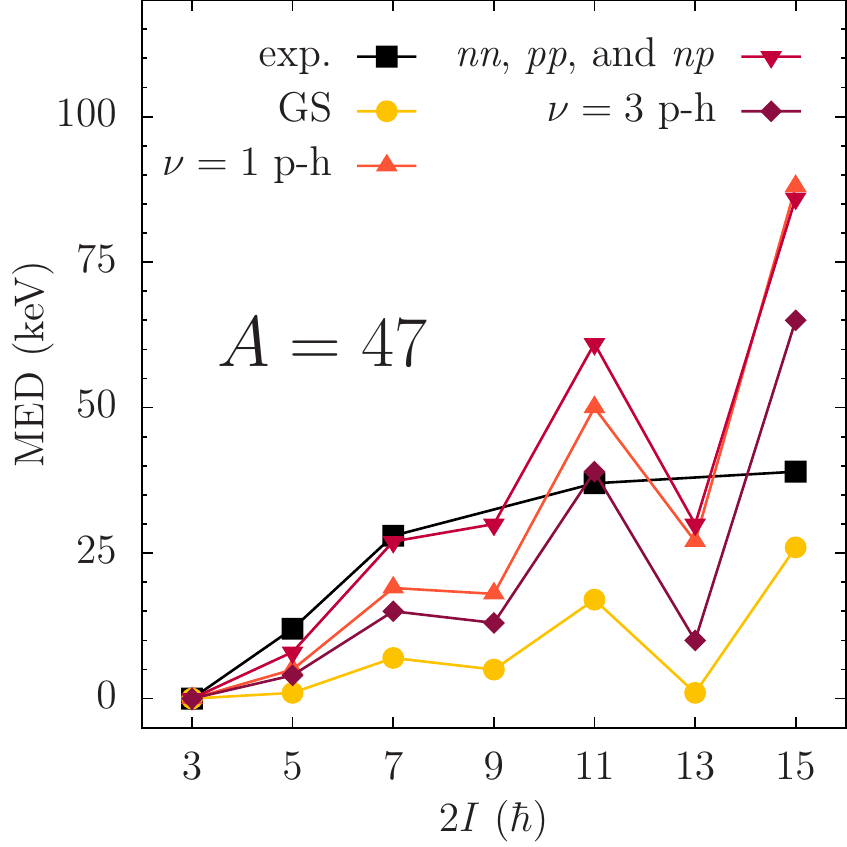}
\caption{(Color online)  
MED versus spin $I$ in the $A=47$ mirror doublet. Experimental data are marked with squares~\cite{(Bur07)}. 
Circles, tringles up, and triangles down shows theoretical results involving the GS configuration,
and $\nu =1$ p-h excitations, and $nn$-, $pp$-, $np$-pairing excitations, respectively. Diamonds mark the DFT-NCCI results involving all configurations depicted in Table~\ref{fig:A47CONF}}.
\label{fig:A47MED}
\end{figure}

The results of the DFT-NCCI calculations for $\Delta E(I)$ and MED$(I)$ are depicted  
 in Fig.~\ref{fig:A47E} and Fig.~\ref{fig:A47MED}, respectively. 
 Different theoretical curves  visualize 
 the role of specific groups of configurations on these observables. 
 Let us concentrate first on the $\Delta E(I)$ curve. Projection from the GS configuration   
 strongly overestimates experimental data. Subsequent admixing of configurations of groups 
 1, 2, and 3 systematically improves the description of experimental data and the process
 nicely converges, at least for low spins. The fact that
 our calculations are free from adjustable parameters allows us to conclude that  our final DFT-NCCI 
 result agrees well with the data for spins up to $I\approx 21/2$ with the exception
 of the lowest two $I=5/2$ and $I=7/2$ states.   In the calculations these states are 
 rotational-like at variance with the experimental data where they are quasi-degenerate.  
 At high spins, above $I = 21/2$, the deviation between theory and experiment 
grows as a function of $I$ up to a band termination at $I=31/2$ where the 
present calculations overestimate the experiment  by 3.180\,MeV.  
This is due to the increasing role of high-seniority multi-broken-pair configurations, which are not 
included in the configuration space. In particular, the structure of  the terminating 
state is dominated by a unique, fully aligned mean-field configuration, see~\cite{(Zdu05),(Zdu07a)}.
Although we focus on the low-spin part of the spectrum, we performed a 
 test calculation which showed that the excitation energy of the $I=31/2$ state projected from the fully
aligned configuration is 9.449\,MeV relative to the $I=3/2$ state projected from the GS configuration  
i.e. only half a MeV below the experimental excitation energy at 10.018\,MeV. It rises hopes 
that also the high-spin part of the spectrum can be reliably well described using the 
DFT-NCCI technique.  

Let us now turn an attention to MEDs.  The results are shown in 
Fig.~\ref{fig:A47MED}. As before, different curves represent the results 
obtained after adding sequentially, atop of the GS configuration, the three groups of 
configurations discussed above. One sees that the single GS configuration leads to MEDs
that are positive but very small, well below the experimental values.  
Inclusion of $\nu =1$ p-h excitations of the first group strongly increases the calculated 
 ${\rm MED}$s,  bringing them very close to experimental values. The addition of pairing-like configurations 
 of the second group does not influence the calculated MEDs. The configurations belonging to 
 the third group partly counterbalance the effect of group one and decrease the MEDs.  Closer inspection shows that the lowering effect is due to the first four configurations belonging to group 3, which 
 are the lowest p-h excitations in the even-particle-number subsystem. The impact of the 
 remaining five configurations belonging to this group is almost negligible.  
 A similar increase (decrease) of theoretical MEDs due to the group 1 (group 3) particle-hole configurations 
and almost negligible effect due to pairing-like configurations  of group 2 was also obtained in the 
calculations performed for the $^{79}$Zr/$^{79}$Y mirror pair in Ref.~\cite{(Lle20)}.

The prerequisite of MEDs are CSB interactions in the nuclear Hamiltonian. Our calculations show, however,  that the net effect is strongly dependent on configuration mixing, which is a very subtle effect  that depends on fine details of the underlying NN interaction and the many-body methods used to describe 
the structure of mirror nuclei under consideration. This explains why MEDs are so difficult to compute accurately. 

\begin{figure}[thb]
\centering
\includegraphics[width=\columnwidth]{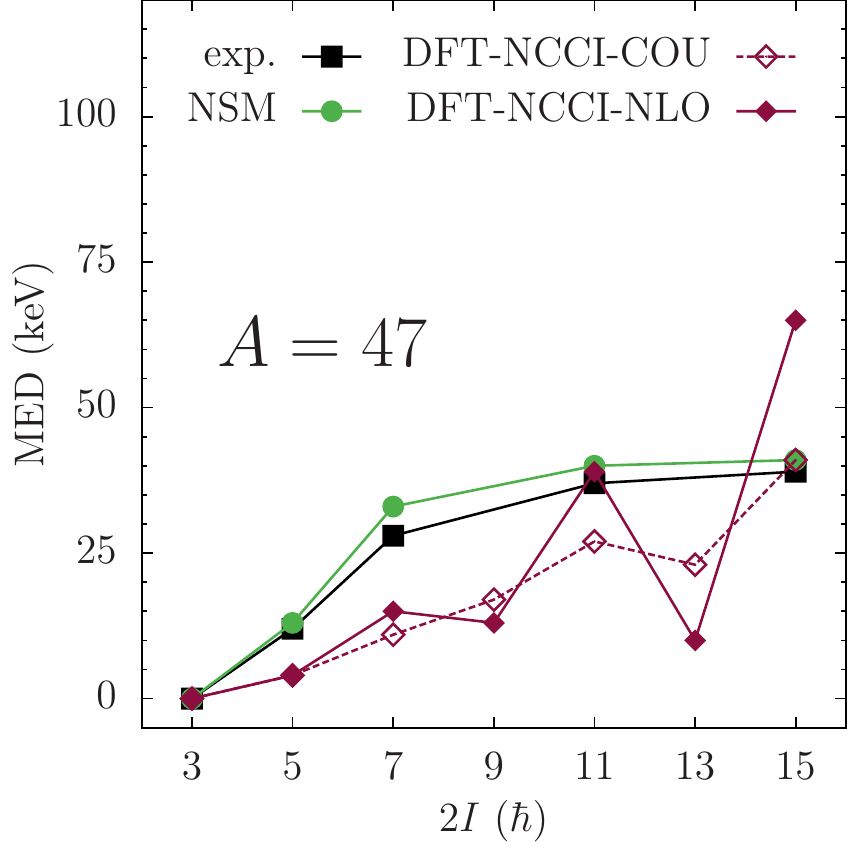}
\caption{
(Color online) Comparison between experimental (filled squares)~\cite{(Bur07)}, shell model (filled circles)~\cite{(Ben15)}, and
two variants of the DFT-NCCI calculations (diamonds) MEDs in the $A=47$ doublet.
Filled diamonds label DFT-NCCI results that include the NLO CSB force. Open diamonds mark the
DFT-NCCI results that use Coulomb interaction as the only source of ISB.
}   
\label{fig:A47MED_SM}
\end{figure}

In the case of the $A=$47 isospin doublet our DFT-NCCI-NLO calculations are in moderate agreement with experiment. They are smaller, approximately by a factor of two, than experimental values for $I < 9/2$. Let us recall, however, that these two lowest-spin states are not well reproduced by our model.  For higher spins $I=11/2$ and $I=15/2$ we slightly overestimate the data.  Calculated MEDs exhibit also a signature staggering with 
MEDs corresponding to energetically favored (unfavored) signature states giving larger (smaller)
MEDs but the effect cannot be verified experimentally using currently available data.  The DFT-NCCI-COU
results are slightly better as compared to the DFT-NCCI-NLO as seen in Fig.~\ref{fig:A47MED_SM}. 
They show much weaker signature staggering and almost perfectly match the data for $I$=15/2.
The DFT-NCCI model is much worse than the NSM, which almost perfectly reproduces low-spin data, see Fig.~\ref{fig:A47MED_SM}.

\subsection{MEDs in the A=43 mirror pair}\label{sec:A43}

\begin{figure}[h!]
\centering
\caption{(Color online)  
Similar to Fig.~\ref{fig:A47CONF} but for the $A=43$ mirror pair.}
\label{fig:A43CONF}
\includegraphics[width=\columnwidth]{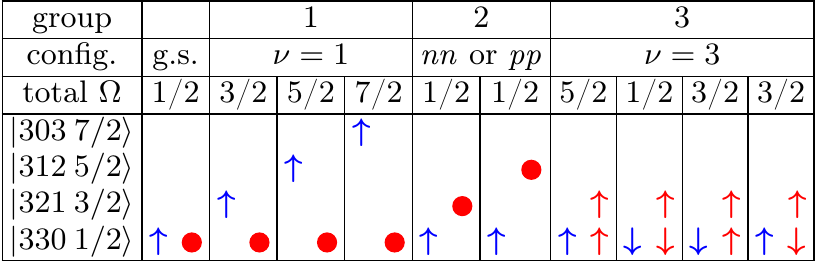}
\end{figure}

Guided by the results for $A$=47 we restrict the configuration space to ten HF solutions, which are 
depicted schematically in Table~\ref{fig:A43CONF}. As well as  the ground-state 
we include three  (all possible) p-h seniority-one ($\nu =1$) excitations in the {\it active\/} space of Nilsson levels. In the second group we admit only two $nn$- or $pp$-pairing $\nu =1$ configurations.  
In the third group, we include $\nu =3$ broken-pair configurations limiting ourselves to the four lowest configurations of this type as shown in the table.  

The $A$=43 mirror nuclei $^{43}$Sc/$^{43}$Ti are typical NSM nuclei. With only three particles outside
the $N=Z=20$ core the collectivity is weak and the spectrum shows rather irregular behavior. Such nuclei
are difficult to reproduce using theoretical techniques based on symmetry-restored mean-field. In this case our 
DFT-NCCI calculations reproduce properly the ground state's spin $I_{\rm GS}$=7/2 but systematically
underestimate excitation energies of higher-spin $I > I_{\rm GS}$ yrast states. With increasing mass and, in turn, increased collectivity of the analyzed mirror pair the agreement systematically improves.

\begin{figure}[thb]
\centering
\includegraphics[width=\columnwidth]{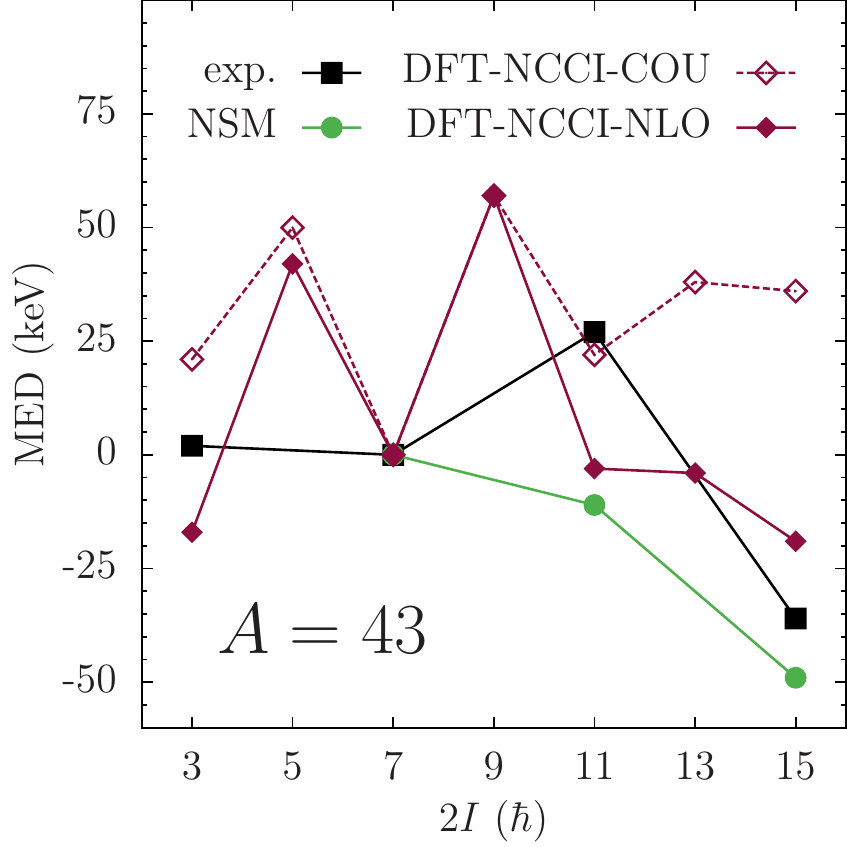}
\caption{
(Color online) Similar to Fig.~\ref{fig:A47MED_SM} but for the $^{43}$Sc/$^{43}$Ti mirror pair. Experimental data were taken from~\cite{(Sin15)} and NSM calculations come from Ref.~\cite{(Ben15)}.}   
\label{fig:A43MED_SM}
\end{figure}

The calculated MEDs are shown in Fig.~\ref{fig:A43MED_SM}. The figure compares two variants 
of our calculations DFT-NCCI-NLO and DFT-NCCI-COU to the NSM results and experimental data
quoted in a review article of Bentley {\it et al.\/}~\cite{(Ben15)}. The DFT-NCCI results were 
obtained using configurations that are schematically depicted in Table~\ref{fig:A43CONF}. What is striking in this
case is the strong influence of the contact NLO CSB force on the calculated MEDs. While
the MEDs  calculated using the DFT-NCCI-COU variant completely disagree with experiment, the 
DFT-NCCI-NLO  results are in reasonable agreement with the data, comparable  (within the considered range of spins) to the NSM of Ref.~\cite{(Ben15)}.

\subsection{MEDs in the A=45 mirror pair}\label{sec:A45}

\begin{figure}[thb]
\centering
\includegraphics[width=\columnwidth]{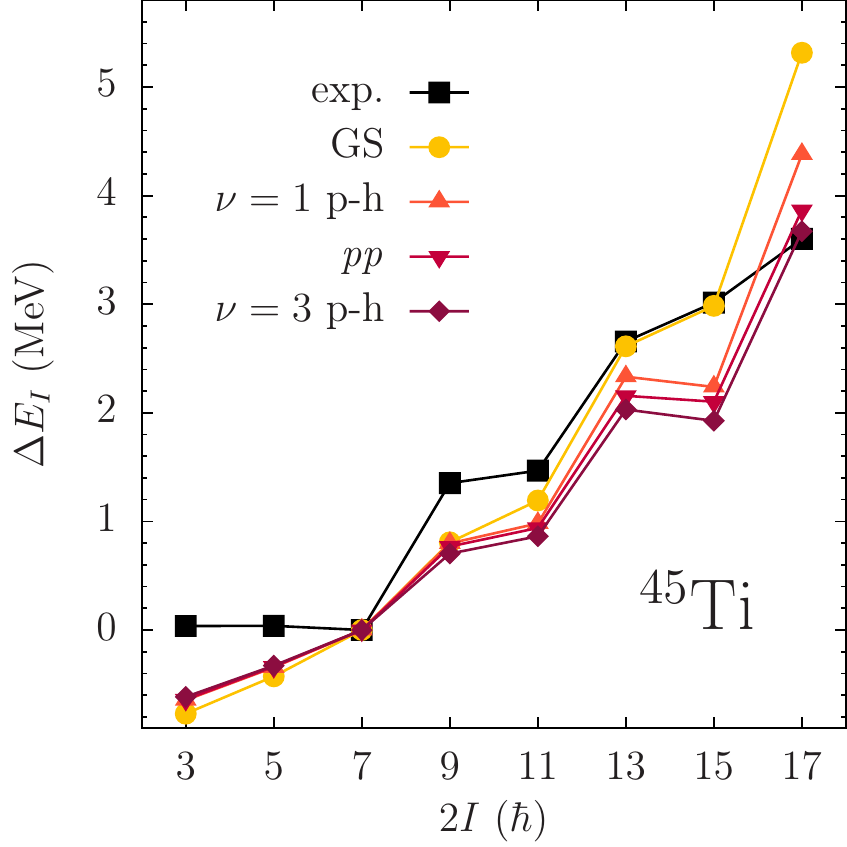}
\caption{(Color online)  
Similar to Fig.~\ref{fig:A47MED_SM} but for the $^{45}$Ti/$^{45}$V mirror pair. Experimental data were taken from~\cite{(Bur08)}.} 
\label{fig:A45E}
\end{figure}

The yrast spectrum in  the $^{45}$Ti/$^{45}$V nuclei shows a very irregular pattern. The three lowest states 
$I=3/2, 5/2$, and 7/2 are nearly degenerated with the $I_{\rm GS} =7/2$ state being the 
ground-state, see Fig.~\ref{fig:A45E}. Higher spin states, on the other hand,  form characteristic 
close-lying doublets that include a pair of $I=9/2, 11/2$ states and a pair of $I=13/2, 15/2$ states in the spin range of interest.   

\begin{figure}
\centering
\caption{(Color online)  
Similar to Fig.~\ref{fig:A47CONF} but for the $A=45$ mirror pair.}
\label{fig:A45CONF}
\includegraphics[width=\columnwidth]{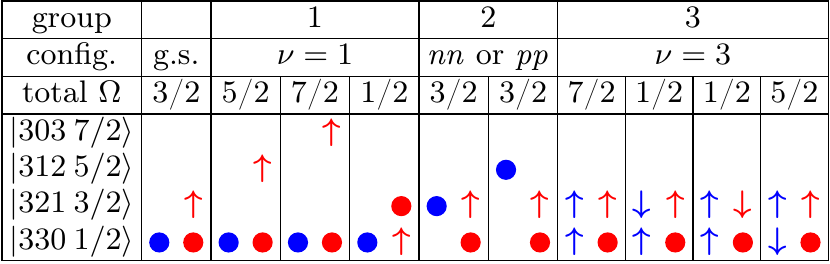}
\end{figure}

The DFT-NCCI calculations only qualitatively reproduce the low spin data as shown in  Fig.~\ref{fig:A45E}.
The presented calculations include 10 configurations which are schematically depicted in Table~\ref{fig:A45CONF}. 
As well as the GS,  
we take into account three $\nu=1$ p-h configurations of group one ($|3/2\rangle\rightarrow |5/2\rangle$, $|3/2\rangle\rightarrow |7/2\rangle$, and
$|\overline{1/2}\rangle\rightarrow |\overline{3/2}\rangle$),  the two lowest $nn$- or $pp$-pairing excitations,  and 
the four lowest seniority-three  ($\nu = 3$)  configurations obtained by breaking a  
($1/2,\overline{1/2}$) pair in the even subsystem. 

The DFT-NCCI calculation fail to reproduce the lowest spin states $I=3/2, 5/2$, and 7/2. 
At variance with the experiment, the calculated spectrum for these states resembles a rotational-like structure 
built upon the deformed $|321\, 3/2\rangle$ Nilsson GS configuration  with the lowest state 
corresponding to $I=3/2$.  Our calculations reproduce, however, quite well the formation of $I=9/2, 11/2$  and $I=13/2, 15/2$ doublets. Their excitation energy is underestimated, but this is an effect of 
normalization to the $I=7/2$ state which is underbound in the theory.

\begin{figure}[thb]
\centering
\includegraphics[width=\columnwidth]{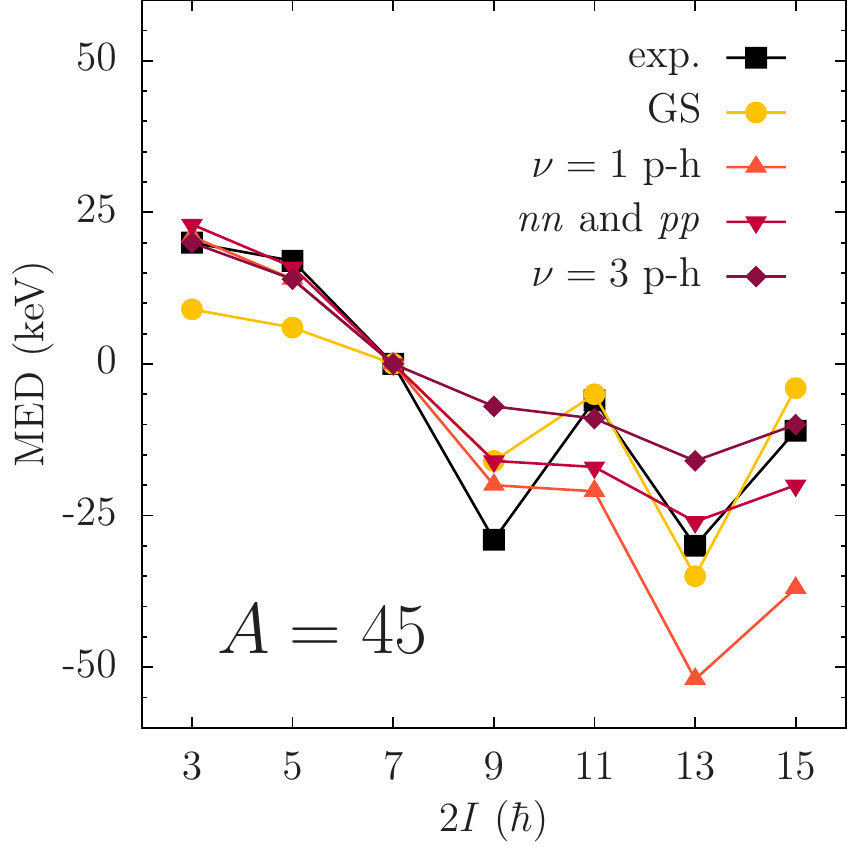}
\caption{(Color online)  
Similar to Fig.~\ref{fig:A47MED} but for the $^{45}$Ti/$^{45}$V mirror doublet. Experimental data were taken from~\cite{(Bur08)}.}
\label{fig:A45MED}
\end{figure}

As shown in Fig.~\ref{fig:A45MED}, empirical MEDs are
positive (negative) for $I< 7/2$  ($I> 7/2$), respectively. For 
$I> 7/2$ we observe strong staggering with large negative MEDs corresponding to $I=9/2$ and $I=13/2$ and negative, albeit much smaller, values for $I=11/2$ and $I=15/2$. This pattern is distinctively 
different as compared to the other cases studied here.  Fig.~\ref{fig:A45MED} shows also the results of
our DFT-NCCI-NLO calculations. Different theoretical curves represent the results that exemplify a role
of different groups of configurations included in the configuration space. It is rewarding to observe that projection from the single GS configuration leads to an excellent agreement with experimental data.  
Admixture of the configurations belonging to the first group improves (deteriorates) the agreement 
for $I<7/2$ ($I>7/2$) states, respectively.  Admixture of the remaining configurations does not affect MEDs for
$I< 7/2$, which are very well reproduced. For $I>7/2$ our final result agrees well with experimental data. 
We reproduce very well experimental MEDs for 
$I=11/2$ and 15/2 but fail to reproduce the staggering, which is too small. In turn, theoretical
results underestimate experimental MEDs for  $I=9/2$ and 13/2.

\begin{figure}[thb]
\centering
\includegraphics[width=\columnwidth]{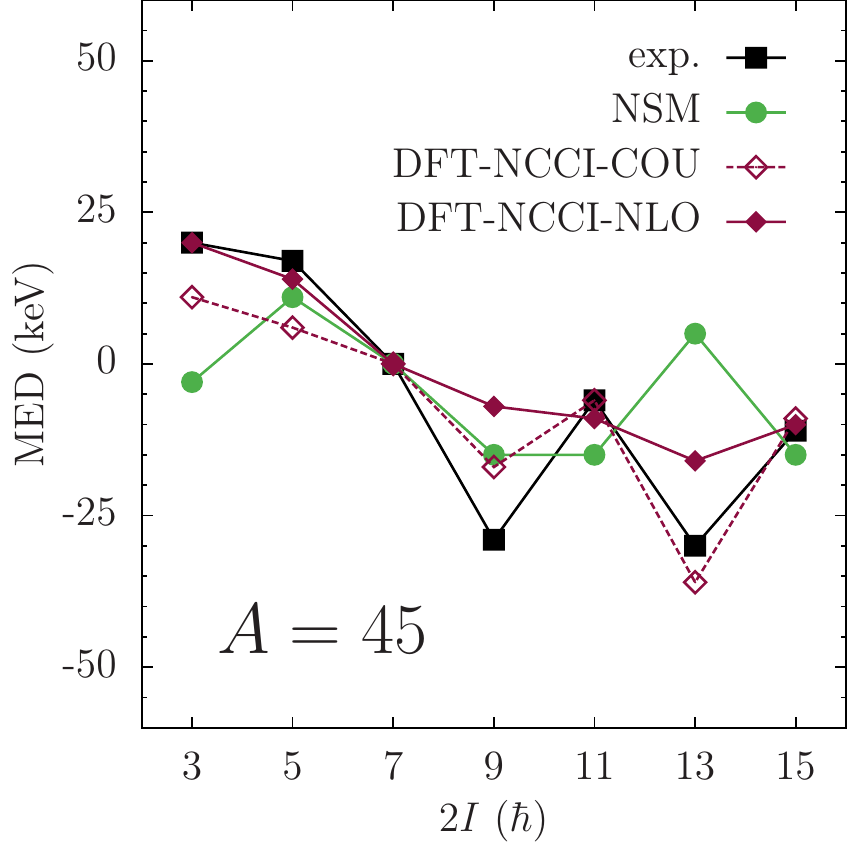}
\caption{  
(Color online) Similar to Fig.~\ref{fig:A47MED_SM} but for the $^{45}$Ti/$^{45}$V mirror pair. Experimental data were taken from~\cite{(Bur08)} and NSM calculations come from Ref.~\cite{(Ben15)}.}   
\label{fig:A45MED_SM}
\end{figure}

The results of our calculations are summarized in Fig.~\ref{fig:A45MED_SM}. 
In the figure we present two variants of the DFT-NCCI results. The results that 
include  the NLO CSB force (filled diamonds)  are compared to the results obtained using 
Coulomb interaction as the only source of ISB (open diamonds). The latter model captures very well
the staggering pattern which is evidently dumped by the short-range CSB force, a property 
that might be used in the future to better constrain the NLO contact CSB force.  Note also 
that in this case our DFT-NCCI results agree well with experimental data and are superior 
to the NSM calculations of Ref.~\cite{(Ben15)}.

\subsection{MEDs in the A=49 mirror pair}\label{sec:A49}

\begin{figure}
\centering
\caption{(Color online)  
Similar to Fig.~\ref{fig:A47CONF} but for the $A=49$ mirror pair. }
\label{fig:A49CONF}
\includegraphics[width=\columnwidth]{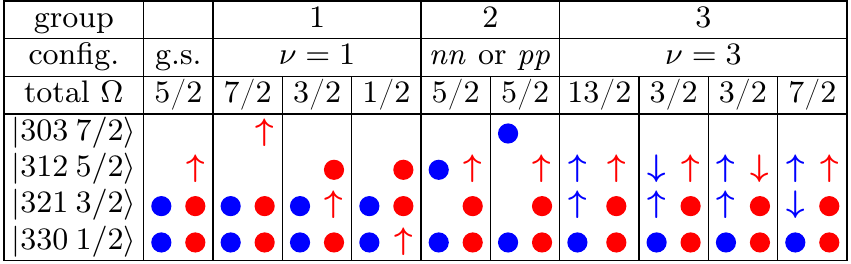}
\end{figure}

Let us finally discuss the results for $A=49$ mirrors. The configurations used 
in the DFT-NCCI calculations are presented in Table~\ref{fig:A49CONF}. Guided by the  results obtained in
lighter cases we include, apart from the GS
three (all within the active space) $\nu =1$ particle-hole configurations, the two lowest $nn$- or $pp$-pairing 
type configurations and the four lowest  $\nu =3$ particle-hole configurations.  

\begin{figure}[thb]
\centering
\includegraphics[width=\columnwidth]{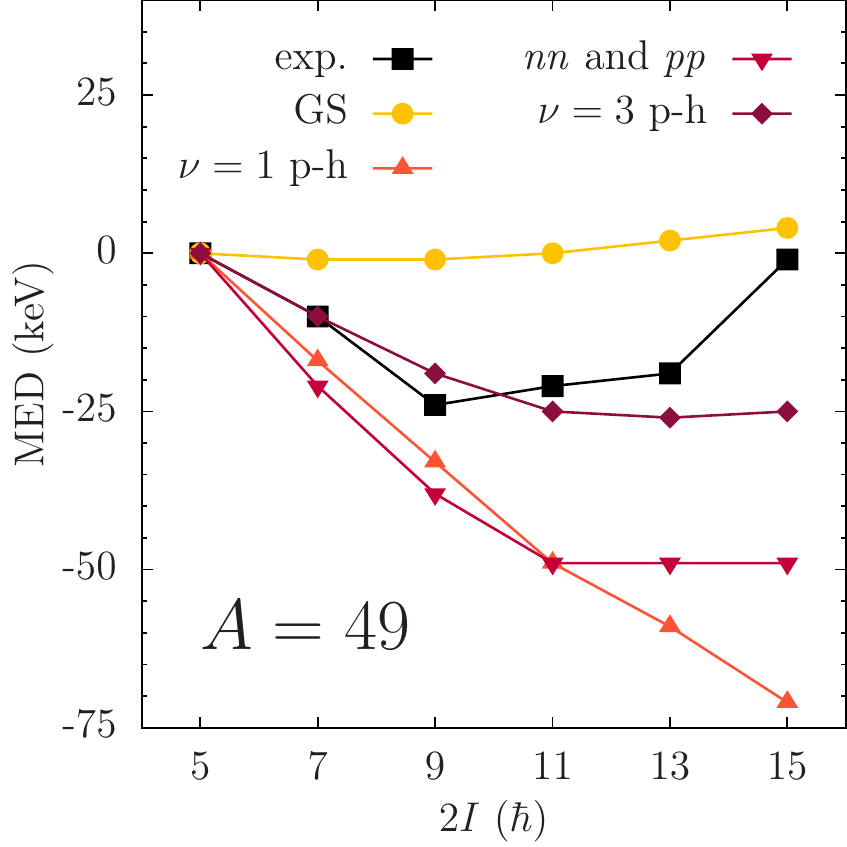}
\caption{(Color online)  
Similar to Fig.~\ref{fig:A47MED} but for the $^{49}$Cr/$^{49}$Mn doublet. Experimental data were taken from~\cite{(Bur08b)}.}
\label{fig:A49MED}
\end{figure}

\begin{figure}[htb]
\centering
\includegraphics[width=\columnwidth]{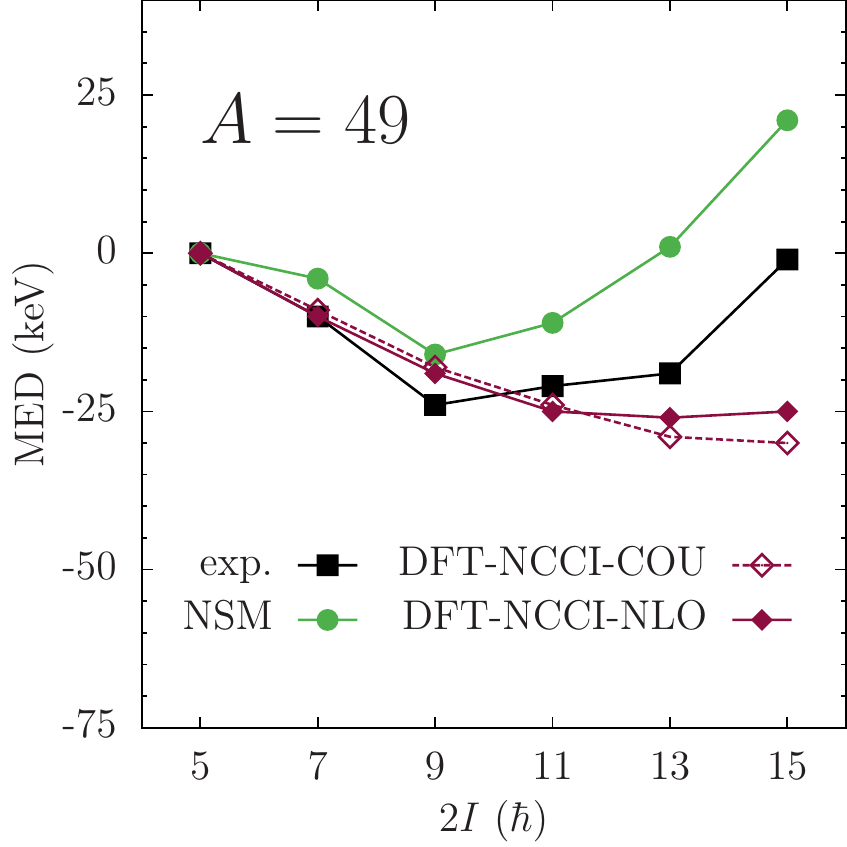}
\caption{(Color online)  Similar to Fig.~\ref{fig:A47MED_SM} but for the $^{49}$Cr/$^{49}$Mn
mirror pair. Experimental data were taken from~\cite{(Bur08b)} and NSM calculations come from Ref.~\cite{(Ben15)}.}   
\label{fig:A49MED_SM}
\end{figure}

The calculated MEDs are shown in Fig.~\ref{fig:A49MED} in comparison with experimental
data taken from Ref.~\cite{(Bur08b)}. Note again that low-spin MEDs are very sensitive 
probes of ISB effects and the underlying nuclear structure. Indeed, only a slight shift in Fermi energy 
when going from the $A=47$ to the $A=49$ mirror pair 
changes completely the behavior of MEDs at low-spins from positive to negative values, respectively. 
It is interesting to observe that our calculations describe low-spin empirical MEDs in $A=49$ mirrors 
very well and account for the change of trend between the $A=47$ and $A=49$ mirrors.
Below $I\leq 13/2$ the level of agreement is similar to the shell-model results, as shown in
Fig.~\ref{fig:A49MED_SM}. The figure shows also that the calculated low-spin MEDs are weakly 
sensitive  to the short-range NLO force.

\begin{figure}[ht!]
\centering
\includegraphics[width=\columnwidth]{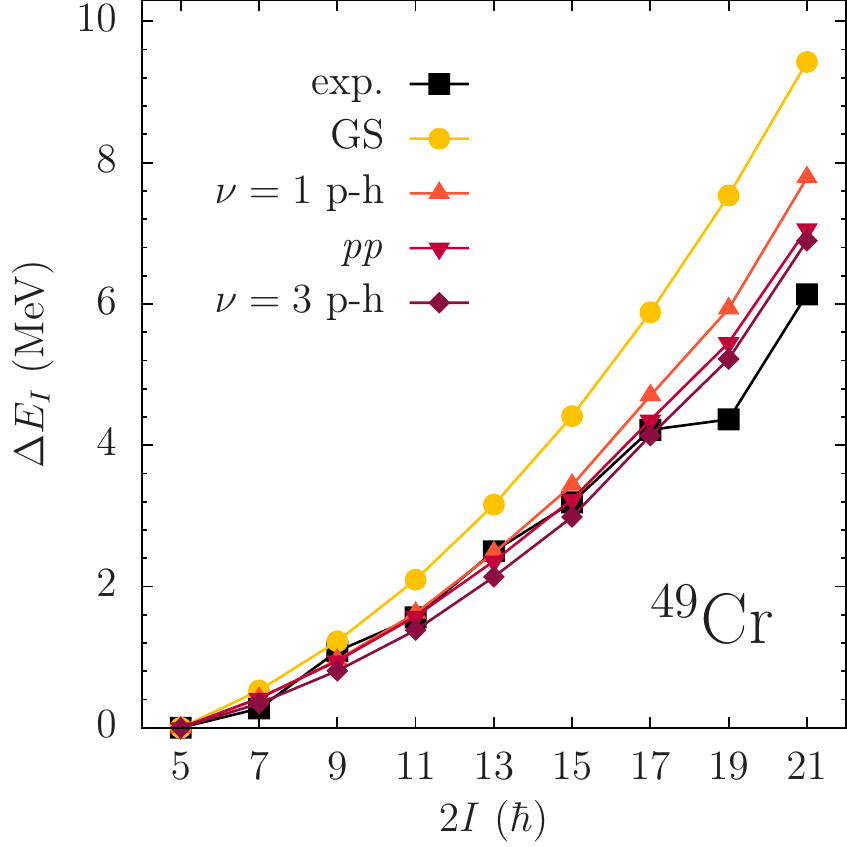}
\caption{(Color online)  
Similar to Fig.~\ref{fig:A47E} but for the $^{49}$Cr/$^{49}$Mn doublet. }
\label{fig:A49E}
\end{figure}

At higher spins, above $I> 13/2$,  the agreement between our calculation and experiment
deteriorates. This is, most likely, due to the restricted configuration space which does not include
higher-lying broken pair configurations.   For the sake of completeness, 
in Fig.~\ref{fig:A49E} we plot also the calculated excitation energy curve $\Delta E(I)$,
which agrees well with experimental data for spins up to $I = 17/2$.

\section{Conclusions}\label{sec:conc}

\begin{figure*}[bht]
\centering
\includegraphics[scale=1.0]{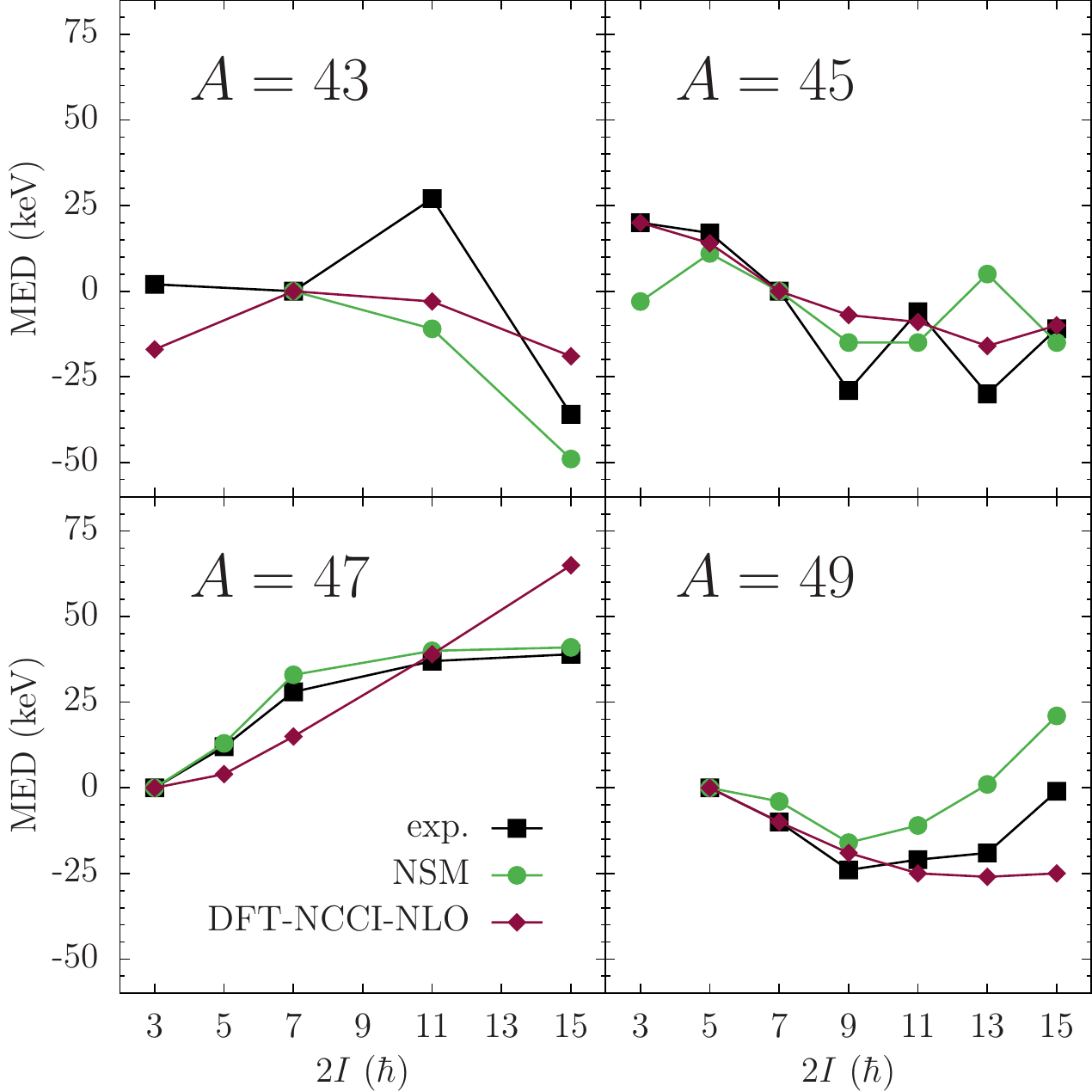}
\caption{(Color online)  
Summary of MED calculations in lower-$fp$ shell $T$=1/2 mirror pairs. Squares represent available experimental data~\cite{(Bur07),(Bur08),(Bur08b),(Sin15)}. Diamonds illustrate  the results of DFT-NCCI-NLO calculations. Shell model results of 
Ref.~\cite{(Ben15)} are marked by circles.} 
\label{fig:ALL}
\end{figure*}

We present a new approach to calculating MEDs, which is based on multi-reference density functional 
theory and involves configuration mixing. The model is applied to MEDs in $A=43$, 45, 47, and 49  $T=1/2$ mirror pairs from the lower $fp$-shell where our calculations can be benchmarked with the 
existing data and NSM results.  The credibility of the DFT-NCCI approach  
to MEDs  is demonstrated in Fig.~\ref{fig:ALL}, which summarizes our results, the existing data, and the 
NSM calculations by Bentley {\it et al.\/}~\cite{(Ben15)}.  Different than before, this time we limit ourselves only to the existing data. The figure 
evidently shows that, for spins $I\leq 15/2$, being the subject of the present study, our model 
is ({\it i\/}) fully capable  of capturing strongly varying experimental trends 
in the function of $A$, and ({\it ii\/}) the accuracy of its predictions is comparable to the NSM.
Let us stress  that our model does not contain adjustable parameters because all its LECs are 
adjusted globally. Moreover, it can be used to all $N\approx Z$ nuclei, in particular 
to  the $A\approx 80$ mass region where the rotational bands are built on very elongated 
shapes, which was demonstrated recently for the case of the $^{79}$Zr/${^{79}}$Y mirror pair in 
Ref.~\cite{(Lle20)}. It should be said that the applicability of the conventional shell model to these 
nuclei is strongly limited due to large model spaces that must involve orbitals originating 
from the $p, f, g, d$ spherical sub-shells.

As well as MEDs, our model also accounts globally, irrespectively of $A$, for the Mirror and 
Triplet Displacement energies in nuclear binding energies, as shown in Refs.~\cite{(Bac18),(Bac19)}. It 
can be also applied to study very subtle effects like isospin impurities~\cite{(Sat09)} or 
isospin-symmetry-breaking corrections to the superallowed $0^+\rightarrow 0^+$ and $T=1/2$ mirror beta decays, see~\cite{(Sat11),(Sat16d), (Kon19)}. This leads us to a general conclusion that 
the DFT-NCCI model is a reliable and internally consistent tool that accounts well
for different observables and pseudo-observables related to isospin symmetry 
violation in $N\approx Z$ nuclei.

Finally, let us formulate conclusions which are specific for the DFT-NCCI applications to MEDs. 
Firstly, our calculations clearly demonstrate that configuration mixing is absolutely 
indispensable.  In order to account quantitatively for MEDs in a low-spin regime one has to include 
seniority one particle-hole 
configurations involving unpaired proton or neutron active Nilsson orbitals
and the lowest seniority-three configurations involving one-broken-pair. These two 
groups of configurations generate opposite contributions to MEDs.
At low spins, $nn$-, $pp$, and $np$-pairing excitations of seniority zero weakly influence
MEDs and, in the first approximation, can be omitted. At present, the calculations are not fully conclusive concerning the 
role of non-coulombic sources of 
isospin symmetry breaking on MEDs, although one has to remember that these terms are 
vital for MDEs and TDEs.  Indeed, they improve (deteriorate) agreement with experiment for
 $^{43}$Ti/${^{43}}$Sc ( $^{45}$Ti/${^{45}}$V), respectively, while  for $A$=47 and 49 the 
 level of agreement with experiment is similar for both the DFT-NCCI-COU and 
DFT-NCCI-NLO variants of the calculations.  In our opinion, MEDs can be used to further optimize
LECs of the contact class III force, in particular, to better constrain $t^{\rm III}_0$ and
$t^{\rm III}_1$ parameters which, in the fit to MDEs, are strongly dependent on each 
other and therefore rather poorly constrained, see Ref.~\cite{(Bac19)} for further details. 
Such a study is in the plans.

\begin{acknowledgments}

This work was supported by the Polish National Science Centre (NCN) under Contracts No 2015/17/N/ST2/04025, 2018/31/B/ST2/02220.
We acknowledge the CIS-IT National Centre for Nuclear Research (NCBJ), Poland for allocation of computational resources.   

\end{acknowledgments}

\bibliographystyle{apsrev4-2}

\bibliography{MED,jacwit34}

\end{document}